\documentclass[doublespacing]{elsart}

\usepackage{amssymb}
\usepackage{graphicx}
\usepackage{amsmath}

\def\beq{\begin{equation}}
\def\eeq{\end{equation}}

\def\rmd{{\rm d}}

\journal{Physics Letters A}

\begin{document}

\begin{frontmatter}

\title{Dixon's extended bodies and impulsive gravitational waves}

\author[iac,icra]{D. Bini},
\ead{binid@icra.it}
\author[ferrara]{P. Fortini},
\ead{fortini@fe.infn.it}
\author[icra,fisica]{A. Geralico},
\ead{geralico@icra.it}
\author[infn]{A. Ortolan}
\ead{ortolan@lnl.infn.it}

\address[iac]
{Istituto per le Applicazioni del Calcolo ``M. Picone'', CNR I-00161 Rome, Italy}

\address[icra]
{International Center for Relativistic Astrophysics - I.C.R.A.\\
University of Rome ``La Sapienza'', I-00185 Rome, Italy}

\address[ferrara]
{Department of Physics, University of Ferrara and INFN Sezione di Ferrara, I-44100 Ferrara, Italy}

\address[fisica]
{Physics Department, University of Rome ``La Sapienza'', I-00185 Rome, Italy}

\address[infn]
{INFN - National Laboratories of Legnaro, I-35020 Legnaro (PD), Italy}

\begin{abstract}
The \lq\lq reaction'' of an extended body to the passage of an exact plane gravitational wave is discussed following Dixon's model.
The analysis performed shows several general features, e.g. even if initially absent, the body acquires a spin induced by the quadrupole structure, the center of mass moves from its initial position, as well as certain \lq\lq spin-flip'' or \lq\lq spin-glitch'' effects which are being observed.
\end{abstract}


\end{frontmatter}

\section{Introduction}

The equations of motion for an extended body in a given gravitational background were deduced
by Dixon \cite{dixon64,dixon69,dixon70,dixon73,dixon74} (hereafter \lq\lq Dixon's model")  in multipole approximation to any order. 
In the quadrupole approximation they read
\begin{eqnarray}
\label{papcoreqs1}
\frac{DP^{\mu}}{\rmd \tau_U}&=&-\frac12R^{\mu}{}_{\nu\alpha\beta}U^{\nu}S^{\alpha\beta}-\frac16J^{\alpha\beta\gamma\delta}R_{\alpha\beta\gamma\delta}{}^{;\,\mu}
\equiv F^{\rm (spin)}{}^{\mu}+F^{\rm (quad)}{}^{\mu}\ , \\
\label{papcoreqs2}
\frac{DS^{\mu\nu}}{\rmd \tau_U}&=&2P^{[\mu}U^{\nu]}+\frac43J^{\alpha\beta\gamma[\mu}R^{\nu]}{}_{\gamma\alpha\beta}
\equiv2P^{[\mu}U^{\nu]}+D^{\rm (quad)}{}^{\mu\nu}\ ,
\end{eqnarray}
where $P^{\mu}=\mu U_p^\mu$ (with $U_p\cdot U_p=-1$) is the total four-momentum of the particle, and $S^{\mu\nu}$ is a (antisymmetric) spin tensor; 
$U$ is the timelike unit tangent vector of the \lq\lq center of mass line'' ${\mathcal C}_U$ used to make the multipole reduction, parametrized by the proper time $\tau_U$.
The tensor $J^{\alpha\beta\gamma\delta}$ is the quadrupole moment of the stress-energy tensor of the body, and has the same algebraic symmetries as the Riemann tensor. 

In this paper we limit our considerations to Dixon's model under the further simplifying assumption \cite{taub64,ehlers77} that the only contribution to the complete quadrupole moment $J^{\alpha\beta\gamma\delta}$ stems from the mass quadrupole moment $Q^{\alpha\beta}$, i.e. we write
\beq
\label{Jdef}
J^{\alpha\beta\gamma\delta}=-3U_p^{[\alpha}Q^{\beta][\gamma}U_p^{\delta]}\ ,\qquad Q^{\alpha\beta}U_p{}_\beta=0\ .
\eeq

In order the model to be mathematically consistent the following additional condition should be imposed \cite{dixon64} to the spin tensor
\beq
\label{Tconds}
S^{\mu\nu}U_p{}_\nu=0\ ,
\eeq
to ensure the correct definition of the various multipolar terms.
It is also convenient to introduce the  spin vector by spatial (with respect to $U_p$) duality
\beq
\label{spinvec}
S^\beta={\textstyle\frac12} \eta_\alpha{}^{\beta\gamma\delta}U_p^\alpha S_{\gamma\delta}\ ,
\eeq
where $\eta_{\alpha\beta\gamma\delta}=\sqrt{-g} \epsilon_{\alpha\beta\gamma\delta}$ is the unit volume 4-form and $\epsilon_{\alpha\beta\gamma\delta}$ ($\epsilon_{0123}=1$) is the Levi-Civita alternating symbol, 
as well as the scalar invariant
\beq
\label{sinv}
s^2=\frac12 S_{\mu\nu}S^{\mu\nu}=S_\mu S^\mu\ , 
\eeq
which is in general not constant along the path.

Within this scheme, a general relation between $U$ and $U_p$ can be obtained by using Eqs. (\ref{papcoreqs1}), (\ref{papcoreqs2}) and (\ref{Tconds}) (see e.g. Eq. (2.17) of Ref. \cite{ehlers77})
\beq
\label{relaUUP}
\left[\mu^2+\frac14R_{\lambda\mu\nu\rho}S^{\lambda\mu}S^{\nu\rho}\right](U^\sigma-u^\sigma)=S^{\sigma\lambda}\left[F^{\rm (quad)}{}_{\lambda}+\frac12R_{\lambda\mu\nu\rho}u^{\mu}S^{\nu\rho}\right]\ , 
\eeq
where
\beq
u^\sigma=-\frac{(U\cdot P)}{\mu^2}P^\sigma+\frac1{\mu^2}P_\lambda D^{\rm (quad)}{}^{\lambda\sigma}\ .
\eeq
In the case of vanishing quadrupole tensor (i.e. $F^{\rm (quad)}{}^{\mu}=0=D^{\rm (quad)}{}^{\mu\nu}$) Eq. (\ref{relaUUP}) reduces to the following one formerly discussed by Tod, de Felice and Calvani (see Eq. (14) of Ref. \cite{tod})
\beq
\label{relaUUPtod}
U^\sigma=-\frac{(U\cdot P)}{\mu^2}\left[P^\sigma+\frac1{2\mu^2\delta}S^{\sigma\lambda}R_{\lambda\mu\nu\rho}P^{\mu}S^{\nu\rho}\right]\ , 
\eeq
where
\beq
\delta=1+\frac1{4\mu^2}R_{\lambda\mu\nu\rho}S^{\lambda\mu}S^{\nu\rho}\ .
\eeq

In Dixon's model there are no evolution equations for the quadrupole moments of the body; therefore the system of equations (\ref{papcoreqs1}) and (\ref{papcoreqs2}), even if completed with conditions (\ref{Jdef}) and (\ref{Tconds}), is not self-consistent and one must assume that all unspecified quantities are  known as intrinsic properties of the body under consideration.
Moreover, the test body assumption means that mass, spin and quadrupole moments (independent among each other) must all be small enough not to contribute significantly to the background metric. Otherwise, backreaction must be taken into account.

We investigate here how a small extended body, with its center of mass initially at rest, reacts to the passage of an exact plane gravitational wave, extending previous results \cite{mohsenietal,mohseni2} limited to the spinning structure of the body.

\section{Motion of extended bodies in the spacetime of an exact gravitational plane wave}

Consider the metric of an exact plane gravitational wave propagating along the $z$ direction of a coordinate frame written in the form \cite{landau}
\beq
\label{gwmetric}
\rmd s^2 =\eta_{\alpha\beta}\rmd x^\alpha\rmd x^\beta-H(\rmd t-\rmd z)^2\ , \qquad x^\alpha=(t,z,x,y)\ ,
\eeq
where $\eta_{\alpha\beta}={\rm diag}[-1,1,1,1]$ and
\beq
H=h_1(t-z)xy+\frac12h_2(t-z)(x^2-y^2)\ .
\eeq
with $h_1$ and $h_2$ two arbitrary functions associated with  the  two polarizations of the wave.
Units are chosen here so that $c=1=G$.

Let us introduce the family of static observers (with respect to the chosen coordinate system) with four-velocity $e_{\hat 0}=1/\sqrt{1+H}\partial_t$ and orthonormal adapted spatial frame
\beq
\label{framestaticobs}
e_{\hat 1}=\frac{H}{1+H}\partial_t+\sqrt{1+H}\partial_z\ , \qquad e_{\hat 2}=\partial_x\ , \qquad e_{\hat 3}=\partial_y\ .
\eeq
These observers are geodesic only when located at $x=0=y$.

Let the \lq\lq center of mass line'' be generic, i.e. with timelike unit tangent vector $U$ given by 
\beq
\label{Udef}
U=\gamma(e_{\hat 0}+\nu^{\hat a} e_{\hat a})\ , \qquad \gamma=(1-\nu^2)^{-1/2}\ , \qquad \nu^2=\delta_{\hat a\hat b}\nu^{\hat a}\nu^{\hat b}\ , \qquad a,b=1,2,3\ .
\eeq
Analogously, let the total four-momentum $P=\mu U_p$ also have the general form 
\beq
\label{Updef}
U_p=\gamma_p(e_{\hat 0}+\nu_p^{\hat a} e_{\hat a})\ , \qquad \gamma=(1-\nu_p^2)^{-1/2}\ , \qquad \nu_p^2=\delta_{ab}\nu_p^{\hat a}\nu_p^{\hat b}\ .
\eeq

Let us assume --without loss of generality-- that the center of mass of the body is initially at rest at the origin of the coordinates, i.e. the associated world line $U_0$ has parametric equations
\beq
\label{traietiniz}
t=\tau_{U_0}\ , \quad z(\tau_{U_0})=0\ , \quad x(\tau_{U_0})=0\ , \quad y(\tau_{U_0})=0\ ,
\eeq
where $\tau_{U_0}$ denotes the proper time parameter, and hence the unit tangent vector reduces to $U_0=e_{\hat 0}|_{x=y=z=0}\equiv \partial_t$.
The passage of the wave (\ref{gwmetric}) will modify the kinematical state of the body (according to Dixon's model), due to the coupling between the wave and body's structure parameters. 
In order to avoid backreaction effects the natural length scales associated with the body, i.e. the \lq\lq bare'' mass $\mu_0$, the natural spin length $|S^\alpha|/\mu_0$ and the natural quadrupolar length $(|Q^{\alpha\beta}|/\mu_0)^{1/2}$, must be small enough if compared with certain background scale, say ${\mathcal L}_{GW}$ (this can be associated, in turn, with the polarization functions $h_1$ and $h_2$ having both the dimensions of 1/length$^2$).
Therefore, in solving the whole set of equations (\ref{papcoreqs1})--(\ref{Jdef}), we will neglect terms which are higher order than the first in the spin as well as quadrupole length scales, according to 
\beq
\label{vincexp}
\mu=\mu_0+\tilde \mu\ , \quad U=U_0+\tilde U\ , \quad P=\mu_0U_0+\tilde P\ , \quad S^{\mu\nu}=\tilde S^{\mu\nu}\ , \quad Q^{\mu\nu}=\tilde Q^{\mu\nu}\ ,
\eeq  
the tilde denoting first order quantities which therefore must be evaluated along the unperturbed center of mass line (\ref{traietiniz}). Hereafter, we will simply use the  expression \lq\lq to first order"  to mean that 
spin and quadrupole terms in the various quantities are retained up to the first order only.

The tangent vectors (\ref{Udef}) and (\ref{Updef}) thus become  
\beq
U=\partial_t+\tilde \nu^{\hat a} \partial_{\hat a}\ , \qquad 
U_p=\partial_t+\tilde \nu_p^{\hat a} \partial_{\hat a}\ ,
\eeq
where $\tilde \nu$ and $\tilde \nu_p$ are first order terms according to Eq. (\ref{vincexp}) and hence
$(\gamma,\gamma_p)\simeq 1$ at that order.

The conditions (\ref{Tconds}) on the coordinate components of the spin tensor imply
\beq
\label{Tconds2}
S_{01}=S_{02}=S_{03}=0\ ;
\eeq
the remaining components can then be re-expressed in terms of the frame components (with respect to the static observer, see Eq. (\ref{framestaticobs})) of the spin vector as
\beq
\label{spinhat}
S_{23}=S^{\hat 1}\ , \qquad 
S_{12}=\frac{S^{\hat 3}}{\sqrt{1+H}}\ , \qquad 
S_{13}=\frac{S^{\hat 2}}{\sqrt{1+H}}\ . 
\eeq

Similarly, the conditions (\ref{Jdef})$_2$ on the quadrupole tensor imply
\beq
Q_{00}=Q_{01}=Q_{02}=Q_{03}=0\ .
\eeq

The spin force turns out to be 
\begin{eqnarray}
F^{\rm (spin)}&=&F^{\rm (spin)}{}^{\hat 2}e_{\hat 2}+F^{\rm (spin)}{}^{\hat 3}e_{\hat 3}\nonumber\\
&=&\frac{1}{2(1+H)}\left[(h_1S^{\hat 2}-h_2S^{\hat 3})e_{\hat 2}-(h_1S^{\hat 3}+h_2S^{\hat 2})e_{\hat 3}\right]\ ,
\end{eqnarray}
to first order in the spin quantities.

The quadrupole force turns out to be 
\beq
F^{\rm (quad)}=F^{\rm (quad)}{}^{\hat 0}e_{\hat 0}+F^{\rm (quad)}{}^{\hat 1}e_{\hat 1}=
-\frac{1}{4(1+H)^{3/2}}(\dot h_2f+2\dot h_1Q^{\hat 2\hat 3})[e_{\hat 0}-e_{\hat 1}]\ ,
\eeq
to first order in the quadrupole quantities, where we have introduced the quantity $f=Q^{\hat 2\hat 2}-Q^{\hat 3\hat 3}$ and the overdot means derivative with respect to time. 
 
Eqs. (\ref{papcoreqs1}) thus reduce to the following set
\begin{eqnarray}
\label{eqmoto}
-\frac{\rmd \tilde \mu}{\rmd \tau_U}&=&F^{\rm (quad)}{}^{\hat 0}\ , \qquad 
\mu_0\frac{\rmd \tilde \nu_p^{\hat 1}}{\rmd \tau_U}=F^{\rm (quad)}{}^{\hat 1}\ , \nonumber\\
\mu_0\frac{\rmd \tilde \nu_p^{\hat 2}}{\rmd \tau_U}&=&F^{\rm (spin)}{}^{\hat 2}\ , \qquad
\mu_0\frac{\rmd \tilde \nu_p^{\hat 3}}{\rmd \tau_U}=F^{\rm (spin)}{}^{\hat 3}\ ,
\end{eqnarray}
so that the spin force affects only motion on the wave front, whereas the quadrupole force acts along the direction of propagation of the wave itself.
The integration constants arising from the mass equation is fixed by imposing $\tilde \mu(0)=0$. 
The three integration constants coming from the equations for the linear velocities $\tilde \nu_p^{\hat a}$ must be left indeterminate at this stage; in fact, they will enter the following Eq. (\ref{eqnu}) for $\tilde \nu^{\hat a}$, and will be fixed by requiring that $\tilde \nu^{\hat a}(0)=0$, according to Eq. (\ref{traietiniz}).

Consider then the evolution equations (\ref{papcoreqs2}) for the spin tensor.
By using the supplementary conditions (\ref{Tconds2}) and Eq. (\ref{spinhat}) they give three algebraic relations between the spatial linear velocities $\tilde \nu^{\hat a}$ of $U$ and $\tilde \nu_p^{\hat a}$ of $U_p$ 
\begin{eqnarray}
\label{eqnu}
\tilde \nu^{\hat 1}&=&\tilde \nu_p^{\hat 1}-\frac{1}{2\mu_0}(h_2f+h_1Q^{\hat 2\hat 3})\ , \nonumber\\
\tilde \nu^{\hat 2}&=&\tilde \nu_p^{\hat 2}+\frac{1}{2\mu_0}(h_1Q^{\hat 1\hat 3}+h_2Q^{\hat 1\hat 2})\ , \nonumber\\
\tilde \nu^{\hat 3}&=&\tilde \nu_p^{\hat 3}-\frac{1}{2\mu_0}(h_2Q^{\hat 1\hat 3}-h_1Q^{\hat 1\hat 2})\ , 
\end{eqnarray}
plus three evolution equations for the spatial components $\tilde S^{\hat a}$ of the spin tensor to be integrated together with the initial conditions $\tilde S^{\hat a}(0)=\tilde S^{\hat a}_0$:
\begin{eqnarray}
\label{eqspin}
\dot {\tilde S}^{\hat 1}&=&-\frac12(h_1f-h_2Q^{\hat 2\hat 3})\ , \nonumber\\
\dot {\tilde S}^{\hat 2}&=&\frac12(h_1Q^{\hat 1\hat 2}-h_2Q^{\hat 1\hat 3})\equiv \mu_0(\tilde \nu^{\hat 3}-\tilde \nu_p^{\hat 3})\ , \nonumber\\
\dot {\tilde S}^{\hat 3}&=&-\frac12(h_1Q^{\hat 1\hat 3}+h_2Q^{\hat 1\hat 2})\equiv -\mu_0(\tilde \nu^{\hat 2}-\tilde \nu_p^{\hat 2})\ ,
\end{eqnarray}
Note that the overdot, used to denote derivative with respect to the coordinate time, coincides  in this case with the differentiation with respect to the proper time $\tau_{U_0}$, as indicated in Eqs. (\ref{traietiniz}).
Eqs. (\ref{eqspin}) thus imply that even if initially absent the spinning structure will be acquired by the body during the evolution, due to its quadrupolar structure.
Similarly, for a purely spinning body the spin components are necessary constant.

Summarizing, the whole set of equation is
\beq
\begin{array}{c@{\hspace{1cm}}l}
({\rm Mass})&\quad
\dot {\tilde \mu}=\displaystyle\frac14(\dot h_2f+2\dot h_1Q^{\hat 2\hat 3})\ , \\ 
({\rm Momentum})&\left\{
\begin{array}{l}
 \dot {\tilde \nu}_p^{\hat 1}=\displaystyle\frac{1}{4\mu_0}(\dot h_2f+2\dot h_1Q^{\hat 2\hat 3})\  \\
 \dot {\tilde \nu}_p^{\hat 2}=\displaystyle\frac{1}{2\mu_0}(h_1S^{\hat 2}-h_2S^{\hat 3})\  \\
 \dot {\tilde \nu}_p^{\hat 3}=-\displaystyle\frac{1}{2\mu_0}(h_1S^{\hat 3}+h_2S^{\hat 2})\  \\
\end{array}
\right. \ , \\
({\rm CM})&\left\{
\begin{array}{l}
\tilde \nu^{\hat 1}=\tilde \nu_p^{\hat 1}-\displaystyle\frac{1}{2\mu_0}(h_2f+h_1Q^{\hat 2\hat 3})\  \\
\tilde \nu^{\hat 2}=\tilde \nu_p^{\hat 2}+\displaystyle\frac{1}{2\mu_0}(h_1Q^{\hat 1\hat 3}+h_2Q^{\hat 1\hat 2})\  \\
\tilde \nu^{\hat 3}=\tilde \nu_p^{\hat 3}-\displaystyle\frac{1}{2\mu_0}(h_2Q^{\hat 1\hat 3}-h_1Q^{\hat 1\hat 2})\  \\
\end{array}
\right. \ , \\
({\rm Spin})&\left\{
\begin{array}{l}
\dot {\tilde S}^{\hat 1}=-\displaystyle\frac12(h_1f-h_2Q^{\hat 2\hat 3})\  \\
\dot {\tilde S}^{\hat 2}=\displaystyle\frac12(h_1Q^{\hat 1\hat 2}-h_2Q^{\hat 1\hat 3})\  \\
\dot {\tilde S}^{\hat 3}=-\displaystyle\frac12(h_1Q^{\hat 1\hat 3}+h_2Q^{\hat 1\hat 2})\  \\
\end{array}
\right. \ . \\
\end{array}
\label{quadrofin}
\eeq

The modification to the initial trajectory (\ref{traietiniz}) of the body after the passage of the wave is then obtained by integrating Eqs. 
(\ref{quadrofin}), taking into account that $\tilde \nu^a={\rmd x^a}/{\rmd \tau_U}$ with initial conditions $x^a(0)=0$, once the profile of the polarization functions $h_1$ and $h_2$ is specified.
We will explore below the two complementary cases of single polarization for the wave.

\subsection{Single polarization: case 1 ($h_1\not=0$, $h_2=0$)}

Let us consider the case of single polarization: $h_2=0$.
The whole set of equations (\ref{eqmoto}) and (\ref{eqspin}) reduces to
\begin{eqnarray}
\label{eqsingle}
&&\dot {\tilde \mu}=\displaystyle\frac12\dot h_1Q^{\hat 2\hat 3}\ , \nonumber\\ 
&&\dot {\tilde \nu}_p^{\hat 1}=\displaystyle\frac{1}{2\mu_0}\dot h_1Q^{\hat 2\hat 3}\ , \quad
\dot {\tilde \nu}_p^{\hat 2}=\displaystyle\frac{1}{2\mu_0}h_1S^{\hat 2}\ , \quad
\dot {\tilde \nu}_p^{\hat 3}=-\displaystyle\frac{1}{2\mu_0}h_1S^{\hat 3}\ , \nonumber\\
&&\tilde \nu^{\hat 1}=\tilde \nu_p^{\hat 1}-\displaystyle\frac{1}{2\mu_0}h_1Q^{\hat 2\hat 3}\ , \qquad
\tilde \nu^{\hat 2}=\tilde \nu_p^{\hat 2}+\displaystyle\frac{1}{2\mu_0}h_1Q^{\hat 1\hat 3}\ , \qquad
\tilde \nu^{\hat 3}=\tilde \nu_p^{\hat 3}+\displaystyle\frac{1}{2\mu_0}h_1Q^{\hat 1\hat 2}\ , \nonumber\\
&&\dot {\tilde S}^{\hat 1}=-\displaystyle\frac12h_1f\ , \qquad
\dot {\tilde S}^{\hat 2}=\displaystyle\frac12h_1Q^{\hat 1\hat 2}\ , \qquad
\dot {\tilde S}^{\hat 3}=-\displaystyle\frac12h_1Q^{\hat 1\hat 3}\ .
\end{eqnarray}
In order to explore the physical content of this situation we assume that all the nondiagonal frame components of the quadrupole tensor vanish, the remaining ones being constant.
Such hypotheses, which can be easily released only leading to an increasing computational work,  will allow to find out explicit solutions for all quantities straightforwardly, without changing the physical interpretation. 
Eqs. (\ref{eqsingle}) then imply $\dot {\tilde \mu}=0$ and $\tilde \nu^{\hat a}=\tilde \nu_p^{\hat a}$ and
\begin{eqnarray}
\label{eqsinglediag}
\dot {\tilde \nu}_p^{\hat 1}&=&0\ , \quad
\dot {\tilde \nu}_p^{\hat 2}=\frac{1}{2\mu_0}h_1S^{\hat 2}\ , \quad
\dot {\tilde \nu}_p^{\hat 3}=-\frac{1}{2\mu_0}h_1S^{\hat 3}\ , \nonumber\\
\dot {\tilde S}^{\hat 1}&=&-\frac12h_1f\ , \quad
\dot {\tilde S}^{\hat 2}=0\ , \quad
\dot {\tilde S}^{\hat 3}=0\ ;
\end{eqnarray}
The  solution corresponding to the initial conditions $\tilde \mu(0)=0$, $\tilde \nu^{\hat a}=0$ and $\tilde S^{\hat a}(0)=\tilde S^{\hat a}_0$ implies 
$\tilde \mu=0$ for the mass and
\beq
\tilde \nu_p^{\hat 1}=0=\tilde \nu^{\hat 1},\qquad 
\tilde \nu_p^{\hat 2}=\frac{1}{2\mu_0}H_1S^{\hat 2}_0=\tilde \nu^{\hat 2}\ , \qquad
\tilde \nu_p^{\hat 3}=-\frac{1}{2\mu_0}H_1S^{\hat 3}_0=\tilde \nu^{\hat 3}\ , 
\eeq
for the center of mass world line, while
\beq
\label{eqsinglediagsol}
\tilde S^{\hat 1}=-\frac12H_1f+\tilde S^{\hat 1}_0\ , \qquad
\tilde S^{\hat 2}=\tilde S^{\hat 2}_0\ , \qquad
\tilde S^{\hat 3}=\tilde S^{\hat 3}_0\ ,
\eeq
for the spin, where we have introduced the notation
\beq
H_1(\tau_U)=\int_0^{\tau_U}h_1(\xi)\rmd\xi\ .
\eeq
As a result we see that the center of mass line generally moves, but not in the direction of propagation of the wave.
Moreover, in the case in which the spin is initially absent ($S^{\hat a}_0=0$) the center of mass remains at rest, but the body acquires a varying spin in the direction of propagation of the wave due to its quadrupolar structure. 
It is also interesting to note that the component of the spin $\tilde S^{\hat 1}$ along the direction of propagation of the wave can change its sign if the duration of the wave is long enough, leading to a spin-flip which can be eventually observed.
Similarly, a suitable polarization function such that $H_1(\tau_U)$ vanishes as $\tau_U\to\infty$ guarantees that asymptotically the spin component $\tilde S^{\hat 1}$ goes back to its initial value, an interesting situation which will be sketched in the next section. 

The modified trajectory turns out to be only affected by the spinning structure and is given by
\beq
\label{traietfin}	
t=\tau_U\ , \qquad
z=\ 0, \qquad
x=\frac{1}{2\mu_0}{\mathcal H}_1S^{\hat 2}_0\ , \qquad
y=-\frac{1}{2\mu_0}{\mathcal H}_1S^{\hat 3}_0\ , 
\eeq
where 
\beq
{\mathcal H}_1(\tau_U)=\int_0^{\tau_U}H_1(\xi)\rmd\xi=\int_0^{\tau_U} \rmd \xi \int_0^{\xi }h_1(\eta)\rmd\eta \ .
\eeq
The spatial orbit is then the line 
\beq
y=-\frac{S^{\hat 3}_0}{S^{\hat 2}_0}x\ .
\eeq

\begin{enumerate}

\item {\it Impulsive GPW}:

In this case $h_1(\xi)=A_1\delta(\xi)$, so that 
\beq
H_1(\tau_U)=\frac{A_1}2\ , \qquad
{\mathcal H}_1(\tau_U)=\frac{A_1}2\tau_U\ ,
\eeq
where the factor of 2 comes from integrating delta function over an interval which has 0 as an extreme, as customary.

\item {\it Sandwich wave with finite amplitude}:

In this case $h_1(\xi)=B_1[\theta(\xi)-\theta(\xi-\xi_0)]$, so that 
\beq
H_1(\tau_U)=\left\{
\begin{array}{ll}
B_1\tau_U &\qquad 0<\tau_U<\tau_U^0\\
& \\
B_1\tau_U^0 &\qquad \tau_U>\tau_U^0\ . \\
\end{array}
\right.
\eeq
and
\beq
{\mathcal H}_1(\tau_U)=\left\{
\begin{array}{ll}
\frac12B_1\tau_U^2 &\qquad 0<\tau_U<\tau_U^0\\
& \\
B_1\tau_U^0\tau_U- \frac12B_1(\tau_U^0)^2&\qquad \tau_U>\tau_U^0\ .\\
\end{array}
\right.
\eeq

\end{enumerate}

\subsection{Single polarization: case 2 ($h_1=0$, $h_2\not=0$)}

Let us consider the case of single polarization: $h_1=0$.
The whole set of equations (\ref{eqmoto}) and (\ref{eqspin}) reduces to
\begin{eqnarray}
\label{eqsingle2}
&&\dot {\tilde \mu}=\displaystyle\frac14\dot h_2f\ , \nonumber\\ 
&&\dot {\tilde \nu}_p^{\hat 1}=\displaystyle\frac{1}{4\mu_0}\dot h_2f\ ,  \qquad
\dot {\tilde \nu}_p^{\hat 2}=-\displaystyle\frac{1}{2\mu_0}h_2S^{\hat 3}\ , \qquad
\dot {\tilde \nu}_p^{\hat 3}=-\displaystyle\frac{1}{2\mu_0}h_2S^{\hat 2}\ , \nonumber\\
&&\tilde \nu^{\hat 1}=\tilde \nu_p^{\hat 1}-\displaystyle\frac{1}{2\mu_0}h_2f\ , \qquad
\tilde \nu^{\hat 2}=\tilde \nu_p^{\hat 2}+\displaystyle\frac{1}{2\mu_0}h_2Q^{\hat 1\hat 2}\ , \qquad
\tilde \nu^{\hat 3}=\tilde \nu_p^{\hat 3}-\displaystyle\frac{1}{2\mu_0}h_2Q^{\hat 1\hat 3}\ , \nonumber\\
&&\dot {\tilde S}^{\hat 1}=-\displaystyle\frac12h_2Q^{\hat 2\hat 3}\ , \qquad
\dot {\tilde S}^{\hat 2}=-\displaystyle\frac12h_2Q^{\hat 1\hat 3}\ , \qquad
\dot {\tilde S}^{\hat 3}=-\displaystyle\frac12h_2Q^{\hat 1\hat 2}\ .
\end{eqnarray}
Let us assume again that all the nondiagonal frame components of the quadrupole tensor vanish, the remaining ones being constant.
Eqs. (\ref{eqsingle2}) thus reduce to
\begin{eqnarray}
\label{eqsinglediag2}
\dot {\tilde \mu}&=&\frac14\dot h_2f\ , \quad 
\dot {\tilde \nu}_p^{\hat 1}=\frac{1}{4\mu_0}\dot h_2f\ , \quad
\dot {\tilde \nu}_p^{\hat 2}=-\frac{1}{2\mu_0}h_2S^{\hat 3}\ , \quad
\dot {\tilde \nu}_p^{\hat 3}=-\frac{1}{2\mu_0}h_2S^{\hat 2}\ , \nonumber\\
\tilde \nu^{\hat 1}&=&\tilde \nu_p^{\hat 1}-\frac{1}{2\mu_0}h_2f\ , \quad
\tilde \nu^{\hat 2}=\tilde \nu_p^{\hat 2}\ , \quad
\tilde \nu^{\hat 3}=\tilde \nu_p^{\hat 3}\ , \nonumber\\
\dot {\tilde S}^{\hat 1}&=&0\ , \quad
\dot {\tilde S}^{\hat 2}=0\ , \quad
\dot {\tilde S}^{\hat 3}=0\ ,
\end{eqnarray}
with solution 
\begin{eqnarray}
\label{eqsinglediagsol2}
\tilde \mu&=&\frac14  h_2f\ , \quad 
\tilde \nu_p^{\hat 1}=\frac{1}{4\mu_0}h_2f\ , \quad
\tilde \nu_p^{\hat 2}=-\frac{1}{2\mu_0}H_2S^{\hat 3}_0\ , \quad
\tilde \nu_p^{\hat 3}=-\frac{1}{2\mu_0}H_2S^{\hat 2}_0\ , \nonumber\\
\tilde \nu^{\hat 1}&=&-\frac{1}{4\mu_0}h_2f\ , \quad
\tilde \nu^{\hat 2}=\tilde \nu_p^{\hat 2}\ , \quad
\tilde \nu^{\hat 3}=\tilde \nu_p^{\hat 3}\ , \nonumber\\
\tilde S^{\hat 1}&=&\tilde S^{\hat 1}_0\ , \quad
\tilde S^{\hat 2}=\tilde S^{\hat 2}_0\ , \quad
\tilde S^{\hat 3}=\tilde S^{\hat 3}_0\ ,
\end{eqnarray}
where the initial conditions $\tilde \mu(0)=0$, $\tilde \nu^{\hat a}=0$ and $\tilde S^{\hat a}(0)=\tilde S^{\hat a}_0$ have been imposed and
\beq
H_2(\tau_U)=\int_0^{\tau_U}h_2(\xi)\rmd\xi\ .
\eeq
From the above relations we see  that in this case the center of mass line generally moves, without specific relations with direction of propagation of the wave. Moreover, if the spin is initially absent ($S^{\hat a}_0=0$), the center of mass  moves along the direction of propagation of the wave due to the quadrupole, but the spin of remains zero. 
In addition, the body acquires a varying mass. 

The modified trajectory turns out to be 
\beq
\label{traietfin2}	
t=\tau_U\ , \qquad
z=-\frac{1}{4\mu_0}H_2f, \qquad
x=-\frac{1}{2\mu_0}{\mathcal H}_2S^{\hat 3}_0\ , \qquad
y=-\frac{1}{2\mu_0}{\mathcal H}_2S^{\hat 2}_0\ , 
\eeq
where 
\beq
{\mathcal H}_2(\tau_U)=\int_0^{\tau_U}H_2(\xi)\rmd\xi\ .
\eeq
The spatial orbit is then the line 
\beq
y=\frac{S^{\hat 2}_0}{S^{\hat 3}_0}x\ .
\eeq

The solution for this case can be found similarly to case 1 for both impulsive GPW ($h_2(\xi)=A_2\delta(\xi)$) and sandwich wave with finite amplitude
($h_2(\xi)=B_2[\theta(\xi)-\theta(\xi-\xi_0)]$).

\section{Discussion and concluding remarks}

We have studied how a small extended body, with the center of mass initially at rest, interacts with an incoming exact plane gravitational wave. The body is spinning and also endowed with a quadrupolar structure.
We have discussed its motion  by assuming that it can be described according to Dixon's model and by solving the corresponding set of evolution equations in the case in which the wave has a single polarization, for simplicity.

A number of interesting results have been discussed.
For instance, in general
a) even if initially absent, the body acquires a spin induced by the quadrupole tensor;
b) the center of mass moves from its initial position and the projection of the orbit on the wave front is a straight line, whose inclination depends on the initial spin of the body;
c) special situations may occur in which certain spin components change their magnitude leading to effects (e.g. spin-flip) which can be eventually observed.

This interesting feature recalls the phenomenon of glitches observed in pulsars: a sudden increase in the rotation frequency, often accompanied
by an increase in slow-down rate (see e.g. \cite{pap1,pap2,pap3} and references therein).
Currently, only multiple glitches of the Crab and Vela pulsars have been observed and studied extensively.
Larger glitches in younger pulsars are usually followed by an exponential
recovery or relaxation back toward the pre-glitch frequency,
while for older pulsars and small glitches the jump
tends to be permanent.
The physical mechanism triggering glitches is not well understood yet, even if  
these are commonly thought to be caused by internal processes.

If one models a pulsar by a Dixon's extended body, then
the present analysis shows that a sort of glitch can be generated by the passage of a strong gravitational wave, due to the pulsar quadrupole structure.
In fact, from Eq. (\ref{eqsinglediagsol}) we see that the profile of a polarization function can be suitably selected in order to fit observed glitches and in particular to describe the post-glitch behavior.

We have just considered here a generic extended body, without exploring the possibility that it could actually represent a real pulsar. In fact, the interior structure of a neutron star requires taking into account all the nuclear and hydrodynamical processes. 
This is beyond the scope of our paper.
Furthermore, the observed slow-down of the period of a pulsar is expected to be associated with gravitational wave emission, whereas
we have neglected backreaction effects on the background field. 
This analysis is enough to argue that the phenomena of pulsar glitches are compatible in principle with a pure relativistic model (Dixon's model).

In view of getting in the next year (apparently) an enhanced phase for the interferometric detection of the gravitational waves (for both LIGO and VIRGO the sensitivity should be 5 times increased in comparison with the present one) 
the effects discussed here --in a very simplified form-- constitute an interesting situation  to be further explored.

\end{document}